# Highly Transparent Lead-Free Piezoelectric Haptic Device

Longfei Song,[1,2,4] Sebastjan Glinsek,[1,2] Nagamalleswara Rao Alluri,[1,2] Veronika Kovacova,[1,2] Michele Melchiorr,[3] Alfredo Blazquez Martinez,[1,2,3] Barnik Mandal,[1,2,3] Juliette Cardoletti,[1,2] Emmanuel Defay[1,2]*

1. Luxembourg Institute of Science and Technology, 41 rue du Brill, L-4422 Belvaux, Luxembourg

2. Inter-institutional Research Group University of Luxembourg - LIST on Ferroic Materials, 41 rue du Brill, L-4422 Belvaux, Luxembourg

3. Department of Physics and Materials Science, University of Luxembourg, 41 rue du Brill, L-4422 Belvaux, Luxembourg

4. Present affiliation: College of Intelligent Systems Science and Engineering, Harbin Engineering University, Harbin 150001, China

*Corresponding author: emmanuel.defay@list.lu

Abstract

Acoustic haptic technology adds touch sensations to human-machine interfaces by integrating piezoelectric actuators onto touchscreens. Traditional piezoelectric haptic technologies use opaque lead-containing ceramics that are both toxic and visible. We have developed a highly transparent lead-free piezoelectric haptic device using potassium sodium niobate (KNN) and transparent conductive oxide thin films. The KNN film, grown on glass, exhibits a pure perovskite phase and a dense microstructure. This device achieves up to 80% transmittance, surpassing lead zirconate titanate (PZT) thin films. It generates an acoustic resonance at 16.5 kHz and produces a peak-to-peak displacement of 1.0 μm at 28 V unipolar, making it suitable for surface rendering applications. This demonstrates the potential of transparent lead-free piezoelectric actuators as an effective alternative to conventional PZT haptic actuators.

**Introduction**

Integrating haptic technology at human-machine interfaces has gained substantial momentum, driven by the demand for more immersive and interactive user experiences.[1-5] Piezoelectric haptic technologies are particularly promising due to their capability to provide high-quality and fast-response vibrotactile feedback.[6] Piezoelectric haptic actuators can induce ultrasonic vibrations of a plate, and as a finger moves across it, the rapid compression and decompression of air create a thin air film with a lubricating effect (Figure1a).[7,8] This allows for real-time control of the friction coefficient of the finger, enabling the emulation of artificial textures on touch surfaces.

Most human-machine interfaces currently utilize transparent glass screens, which are ubiquitous in everyday life. However, existing piezoelectric haptic technologies often rely on bulky components and opaque materials, which can compromise the visual integration of the device with its surroundings. Specifically, lead zirconate titanate (PZT) ceramic actuators are commonly mounted beneath these glass screens to produce haptic effects.[9,10] These actuators (typically 500 μm thick), being standalone components, are challenging to miniaturize and assemble, which limits their integration and increases production costs.

A notable trend in the development of haptic actuators is the push towards invisibility. The advent of transparent thin-film actuators offers a transformative solution, enabling the creation of haptic devices that are both functionally effective and visually seamless. Research into transparent piezoelectric haptic actuators predominantly focuses on PZT thin films due to their superior piezoelectric properties and optical transparency.[11-14]

However, the toxicity of lead has led to regulatory constraints under the European Union's 'Restriction of Hazardous Substances' (RoHS) directive, which mandates a reduction in the use of PZT in electronics. Consequently, significant efforts are being directed towards developing lead-free piezoelectric materials as alternatives to PZT.[15-19]

Here, we developed transparent piezoelectric haptic actuators using lead-free potassium sodium niobate (KNN) thin films, chosen for their comparable piezoelectric properties to PZT.[20,21] KNN films, doped with 1% manganese, were deposited via chemical solution process. The transparent actuators consist of a 100 nm indium tin oxide (ITO) layer, a 600 nm KNN layer, and a 200 nm fluoride tin oxide (FTO) layer, integrated on glass (Figure 1b), achieving over 80% transmittance in the visible range. X-ray diffraction and microstructure analyses confirmed the well-crystallized perovskite phase of the KNN film. The haptic device (10×50×0.3 mm³) constructed with four KNN actuators exhibits an acoustic resonance at 16.48 kHz. At 28 V unipolar, the acoustic wave achieves a peak-to-peak displacement of 1 μm, suitable for surface rendering applications.

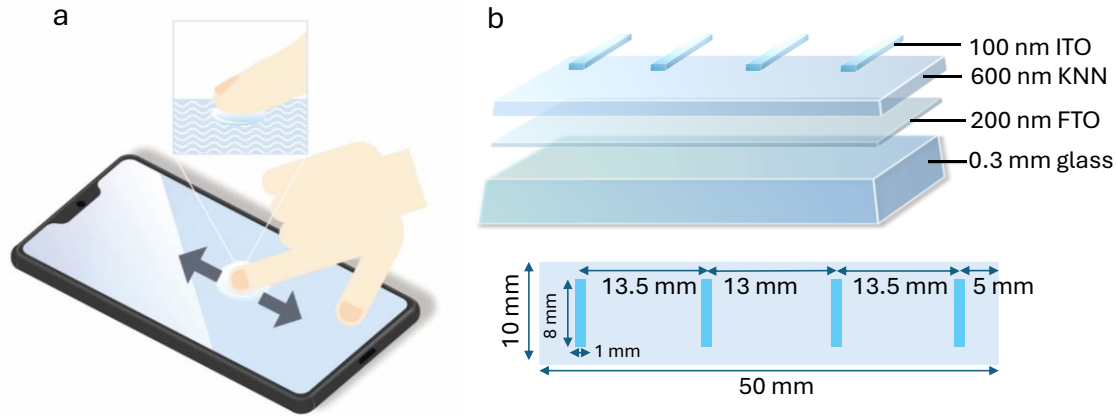

**Figure 1. Schematic illustrations of squeezed air film effect and haptic device**. Schematic illustration of a friction modulation principle based on the squeezed air film effect[7], showing a human finger sliding on an acoustically vibrating plate that is driven by piezoelectric actuators. **b.** Schematic demonstration of lead-free transparent haptic device constructed with a100 nm ITO layer, a 600 nm KNN layer, and a 200 nm FTO layer on glass. The top-view schematic shows the geometry of the device.

**Results**

**Finite element modelling**

3D finite element modelling (FEM) was conducted using COMSOL 5.5 to design the haptic device. In the model, four actuators (1 × 8 × 0.1 mm³) were created on a glass plate (10 × 50 × 0.3 mm³, Figure 1b). Thicker actuator (0.1 mm) was used in the 3D FEM in order to reduce the number of created elements and simplify calculations. It is worth noting that the film thickness does not affect device deflection because firstly, the actuator mass is negligible relative to the glass plate which cannot significantly influence the elastic properties of the device, and secondly, the lateral force generated by e31-mode actuators is voltage-dependent, not thickness-dependent.[22] Electrode influence was ignored due to their negligible thickness. The voltage and ground were directly applied on both sides of the piezoelectric layers. The glass properties used in modelling included a density of 2430 kg m⁻³, a Young's modulus of 74.8

GPa, and a Poisson's ratio of 0.238, corresponding to the SCHOTT AF32 glass used in experiments.[23] The effective transverse piezoelectric coefficient $e_{31,f}$ was set to -3 C m$^{-2}$.

When voltage was applied to the electrodes, the device induced a standing antisymmetric (A0) Lamb wave at a resonance frequency of 17.2 kHz, with four waves and eight nodal lines along the x-axis (Figure 2). Each actuator was precisely positioned at the antinodes, as expected from modelling. At 28V, the peak-to-peak displacement at the antinodes predicted by the modelling was 1μm, confirming the device's design validity.

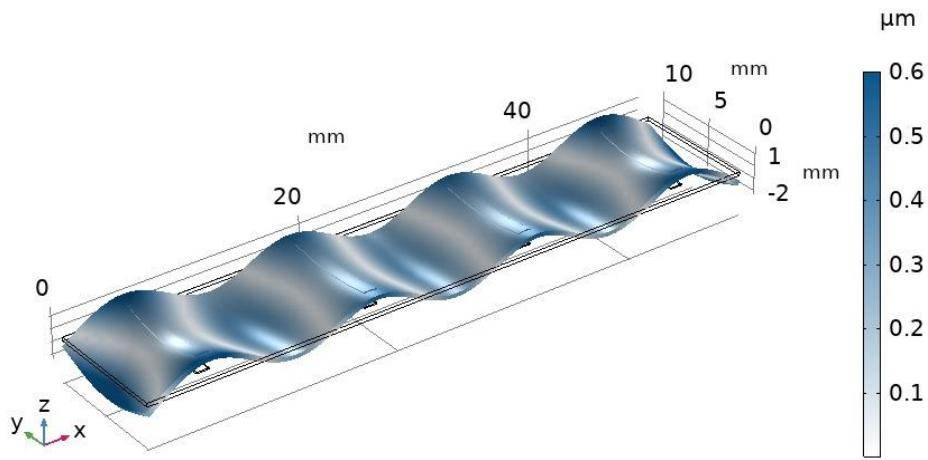

**Figure 2. Finite element modelling.** Mode-shape of haptic device actuated by four piezoelectric actuators at 17.2 kHz and 28 V. The color bar represents the out-of-plane displacement of the device surface.

**Growth, transparency, phase composition and microstructure**

Conductive oxide thin films serve as transparent electrodes, with fluorine-doped tin oxide (FTO) as the bottom electrode and indium tin oxide (ITO) as the top electrode. Both were sputtered in-house. To enhance conductivity and transmittance, ITO was post-annealed at 300 °C in air, and FTO at 600 °C in oxygen, as detailed in Methods. The conductivities are 700 S cm$^{-1}$ for 100 nm-thick ITO and 15 S cm$^{-1}$ for 200 nm-thick FTO.

The 1% Mn doped KNN films were deposited through a standard chemical solution process using 2-methoxyethanol-based solution with acetate precursors,[24,25] as described in Methods. The solution was spin-coated and pyrolyzed at 450 °C, with the deposition-pyrolysis cycle repeated four times to reach a thickness of 200 nm. The film was then crystallized at 680 °C.

This process was repeated multiple times to achieve thicker films. More details are described in Methods. Our previous paper has confirmed that Mn-doping is an effective way to reduce leakage current in KNN thin films.[26] The films grown on platinised silicon have also low losses, can withstand large electric fields and show piezo response ($e_{31,f}$ up to -15 C m$^{-2}$)[26] that is comparable to sputtered films on platinised silicon (-12 C m$^{-2}$)[27].

Thicker films generally exhibit improved piezoelectric properties. However, the 1 μm-thick film cracks due to the strong thermal stress at the film-glass interface. Therefore, a 600 nm film thickness was chosen for the haptic device.

The optical transmittance spectrum of a 1 μm KNN film grown on FTO-coated glass (Figure 3a) exhibits strong oscillations from internal reflections within the KNN multilayer structure. The average transmittance in the visible range (380–700 nm) is 80%, surpassing the values for PZT films. Note that we observed no significant influence of film thickness on transparency, aside from the oscillations arising from the layer-by-layer crystallization process during film growth (supplementary Figure 1). For comparison, Ueda et al. reported average transmittances in the visible range of 70% for sputtered PZT films and 60% for solution-deposited PZT films on FTO glass.[28] The inset of Figure 3a shows the fabricated transparent haptic devices, demonstrating their high transparency.

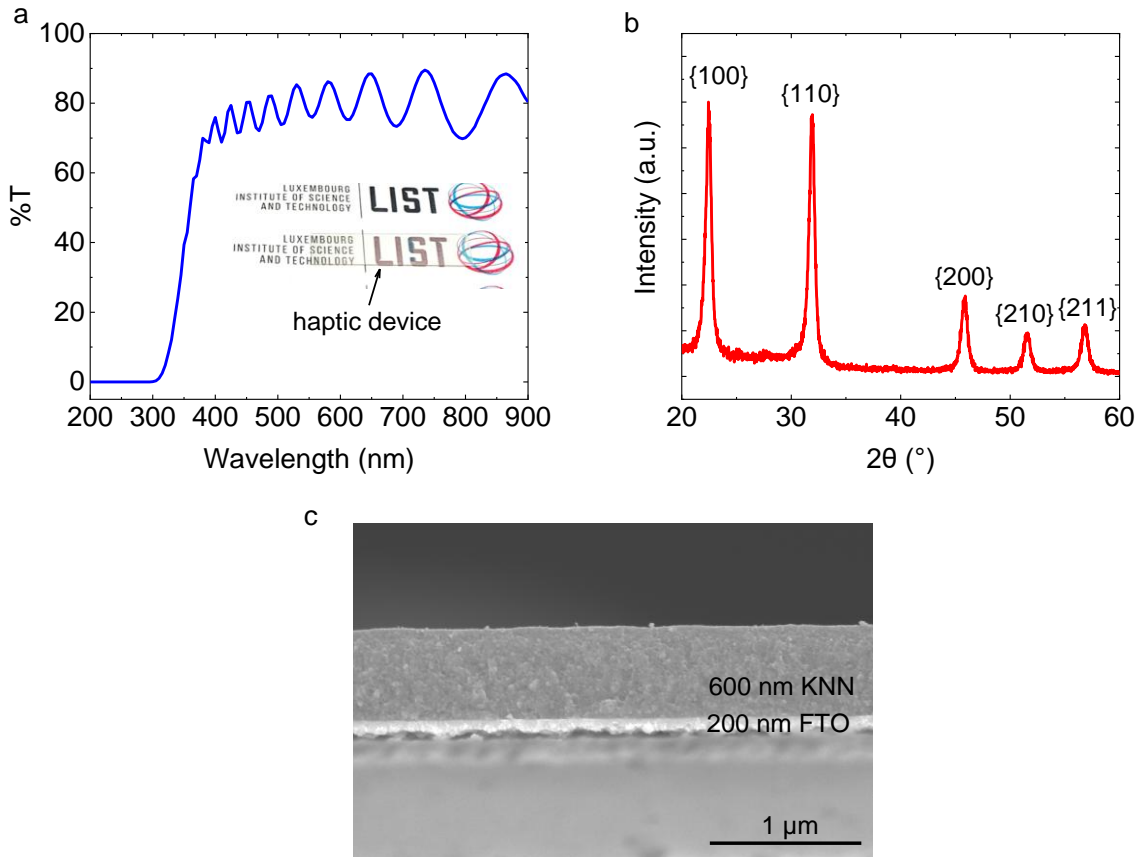

**Figure 3. KNN film's optical and structural characterisation.** a. Optical transmittance spectrum of 1 μm KNN film grown on 200 nm FTO-coated SHOTT AF32 glass. The inset is the optical appearance of haptic devices constructed by 100 nm-ITO/600 nm KNN/200 nm-FTO/0.3 mm-glass substrate. b. GIXRD pattern and c. cross-sectional SEM micrograph of KNN film on FTO grown on glass.

Figure 3b displays the grazing-incidence X-ray diffraction (GIXRD) pattern of the KNN film deposited on FTO glass. The pattern reveals a typical perovskite phase, identified by powder diffraction file No. 00-065-0275,[29] without any secondary phases. The cross-sectional scanning electron microscope (SEM) image (Figure 3c) reveals a dense granular microstructure, which supports the film's well-crystallized nature. Additionally, the interface between the KNN film and the FTO electrode appears clean, indicating that no interdiffusion has occurred.

**Haptic device characterisation**

The haptic device was constructed with four KNN haptic actuators according to the design provided by modelling. SCHOTT AF32 glass, widely used in MEMS and semiconductor industries,[23] was utilized as the resonating plate. The device's geometry and structure are shown in Figure 1b. Detailed information on the experimental setup can be found in the Methods section.

Device performance was assessed by measuring out-of-plane displacement using a Polytech laser doppler vibrometer. During the tests, four actuators were connected in series to reduce the current required, which means that the total applied voltage was four times the voltage applied to each actuator. The out-of-plane displacement at 28 V applied to each actuator (14 V AC + 14 V DC, unipolar) was measured at the anti-node position and is shown as a function of frequency in Figure 4a. Resonance was detected at 16.48 kHz, which aligns well with FEM predictions. The damping loss factor ($\eta$) was determined using the formula $\Delta f / f$, where $\Delta f$ represents the full width at half maximum and f is the resonance frequency. From the frequency sweep shown in Figure 4a, $\eta$ was calculated to be 0.0238. After determining $\eta$, we extracted the effective transverse piezoelectric coefficient $e_{31,f}$ from modelling by matching experimental displacements, yielding -3 C m$^{-2}$. Our previous work validated this method, as the extracted value closely matched the experimentally measured value.[14, 22]

Figure 4b illustrates the peak-to-peak out-of-plane displacement as a function of applied unipolar voltage, showing an increase in displacement with rising voltage. The nonlinear dependence of displacement likely results from charged defects in the film, which pin polarization at low electric fields, as has observed in our previous device made of inkjet printed PZT film.[12]

A 2D surface displacement map, measured at a unipolar voltage of 28 V per actuator and at its resonant frequency, is presented in Figure 4c. The device achieved a peak-to-peak out-of-plane displacement of 1.1 µm, meeting the requirement for haptic rendering (e.g., 1 µm deflection at the resonant frequency)[7]. During the 2D displacement mapping sweep, data were acquired at 1-second step intervals, resulting in a collection of 2000 data points over a period exceeding 30 minutes. The device experienced more than 60 million cycles per hour, and the displacement signal remained constant throughout the experiments.

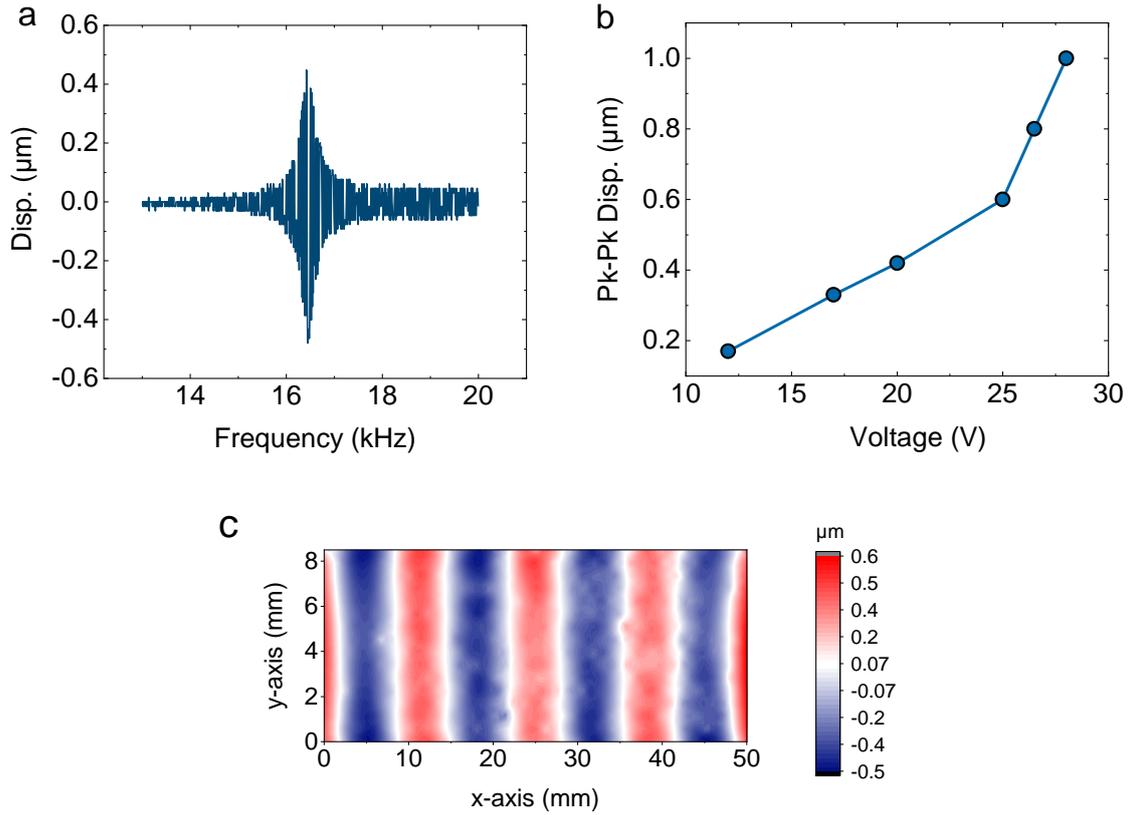

**Figure 4. Haptic device measurements. a.** Out-of-plane displacement as a function of frequency, tested at 28 V unipolar applied to each actuator. **b.** Peak-to-peak displacement as a function of unipolar voltage, measured at 16.48 kHz. **c.** 2D surface displacement mapping of the entire haptic device measured at 28 V unipolar applied to each actuator at 16.48 kHz.

We also evaluated the dielectric properties of the transparent haptic actuator. The dielectric property was measured as functions of DC voltage with a probing AC signal of 0.05V at 1 kHz, using TF Analyzer 2000 (aixACCT), as shown in Supplementary Figure 2. At zero field, relative permittivity ($\varepsilon_r$) and dielectric loss (tan$\delta$) are 400 and 0.15, respectively. The polarization versus electric field (P-E) loop of ITO/KNN/FTO/glass device is shown in supplementary Figure 4, which is much more rounded due to the low conductivity of our in-house FTO electrodes (conductivity 15 S cm$^{-1}$). These electrodes act as series resistors, reducing the voltage applied across the KNN film. Improving FTO conductivity will be explored in future work.

**Discussions**

Our solution-processed Mn-KNN film, characterised in detail in our previous work,[26] demonstrated an e$_{31,f}$ of up to -15 C m$^{-2}$ when deposited on platinised silicon. The permittivity $\varepsilon_r$ can reach up to 900, while the dielectric loss tan$\delta$ is below 0.05 (see supplementary Figure 3).[26] However, the $e_{31,f}$ of films on FTO glass dropped to -3 C m$^{-2}$ with effective loss increased to 0.15 due to the poor conductivity of our in-house FTO (conductivity 15 S cm$^{-1}$), which acted as a series resistor. However, the device deflection exceeds 1 μm, thanks to the structural design including thinner glass plate and incorporation of four actuators to drive the device. Despite the FTO's low conductivity, our transparent KNN actuators perform comparably to PZT-based actuators with similar structures. Ueda's ITO/PZT/FTO/glass actuator achieved an $e_{31,f}$ of −4.5 C m$^{-2}$,[28] slightly higher than our ITO/KNN/FTO actuator.

We also compared our device to transparent haptic devices using PZT thin films. Glinsek et al's transparent e33-mode haptic device, made of 1 μm-thick PZT and interdigitated ITO electrodes, achieved a 1 μm displacement at 60 V and 73 kHz, with 75% visible-range transmittance.[11] Hua et al's e31-mode PZT device, with 2 μm-thick PZT and ITO electrodes (250 nm thick top and 500 nm thick bottom), achieved a 1 μm displacement at 10 V and 22.5 kHz, with 75% transmittance.[13] Despite the FTO's low conductivity, our KNN haptic device produced a 1 μm displacement at 28 V and 16.5 kHz, enabling lower operation voltage compared to Glinsek's e33-mode PZT device. Although Hua's e31-mode PZT device performs better, it requires thicker ITO electrodes to improve conductivity and prevent lead diffusion.

Table 1. Comparison of our device performance with previous lead-based device reported in the literatures.

| Piezoelectric Material | Device structure | Optical transmittance | Voltage per μm in displacement | Literature |
| --- | --- | --- | --- | --- |
| PZT | 100 nm-IDE ITO / 1 μm-PZT / fused silica glass | 75% | 60 V per μm | Glinsek et al.[11] |
| PZT | 250 nm-ITO / 2 μm-PZT / 500 nm-ITO / Corning Lotus™ NXT Glass | 75% | 10 V per μm | Hua et al.[13] |
| KNN | 100 nm-ITO / 600 nm-KNN / 200 nm FTO / SCHOTT AF32 glass | 80% | 28 V per μm | This work |

**Conclusion**

We have successfully grown lead-free piezoelectric KNN films on FTO-coated glass, achieving a pure perovskite phase and a dense microstructure. We developed a lead-free transparent haptic device consisting of ITO/KNN/FTO/glass. This device demonstrated an out-of-plane displacement of 1 μm at 28 V and 16.48 kHz, making it suitable for surface rendering applications. Additionally, the device achieved a high transmittance of up to 80% in the visible range, exceeding the transmittance reported for PZT haptic devices. These findings highlight the potential of lead-free thin films in developing transparent piezoelectric actuators for surface haptic rendering, offering a promising alternative to conventional lead-containing devices.

**Methods**

<u>Solution preparation</u>: A 0.4 M, 50 mL solution of 1 mol% manganese-doped potassium niobate ($K_{0.5}Na_{0.5}Nb_{0.99}Mn_{0.01}O_3$) was synthesized using 2-methoxyethanol-based solution modified with acetylacetone and excess alkali precursors.

The stoichiometric raw materials included potassium acetate ($CH_3OOK$, anhydrous, ⩾99%, Sigma-Aldrich), sodium acetate ($CH_3COONa$, anhydrous, ⩾99%, Sigma-Aldrich), manganese (II) acetate tetrahydrate ($C_4H_6MnO_4 \cdot 4H_2O$, 99.9%, Sigma-Aldrich), niobium (V) ethoxide (($CH_3CH_2O)_5Nb$, 99.99%, Thermo Fisher (Kandel) GmbH), 2-methoxyethanol (2-MOE), acetic acid, acetylacetone. 10% excess potassium acetate and 5% excess sodium acetate were added to compensate for volatilization losses.

First, manganese acetate trihydrate was freeze-dried to an anhydrous state. Afterwards, it was dissolved into acetic acid with magnetic stirring for 1 hour to form a 0.1 M solution. Second, acetate precursors (K, Na, Mn) were dissolved in acetic acid and refluxed at 135°C for 10 min with magnetic stirring under an argon atmosphere. Third, niobium (V) ethoxide was dissolved in 2-methoxyethanol with acetylacetone added dropwise, then combined with the alkali solution and refluxed at 135°C with magnetic stirring for 4 hours. Distillation was done to remove approximately half of solvent. The final solution concentration was adjusted to 0.4 M with 2-methoxyethanol, resulting in a transparent Mn-KNN solution.

<u>KNN films deposition</u>: KNN films were spin-coated at 3000 rpm for 30 seconds, then pyrolyzed at 450°C for 1 minute on a hot plate. After four deposition-pyrolysis cycles, the 200 nm-thick film was crystallized at 680°C in air using a rapid thermal annealing furnace (AS-Master, Annealsys). Thicker films were obtained by repeating this entire process multiple times.

FTO and ITO films deposition: FTO (fluorine-doped tin oxide) and ITO ($In_2O_3/SnO_2$ 90/10 wt%) films were sputtered onto the substrates by using an AJA Orion 8 sputtering tool. The magnetron gun equipped with either 3-inch diameter FTO (99.9+% purity) or ITO (99.998% purity) targets was powered by 300 W RF generator and operated in a non-reactive Ar atmosphere using a constant Ar flow, which was set at 1 mTorr during the sputtering process. In the deposition of FTO and ITO films the sputtering power applied to the target was fixed to 110 and 130 W, respectively.

Film characterisations: Optical transmittance spectrum was recorded using a Perkin Elmer Lambda 1050 UV/Vis spectrophotometer. Film phase composition was analysed with a Bruker D8 Discover XRD using Cu-Kα radiation, and GIXRD patterns were recorded at a 0.5° incidence angle over a 2θ range of 20°–60° with a 0.02° step. Cross-sectional microstructures were examined using an FEI Helios NanoLab 650 SEM, with a 1 nm sputter-coated Pt to prevent charging.

Haptic performance measurement: A 0.6 μm-thick KNN layer was deposited on in-house FTO-coated 5×5 $cm^2$ SHOTT AF32 glass wafer. ITO films were patterned by lithography to create four actuators based on FEM design. The device was cut to the desired dimensions using a wire saw, and copper wires bonded with silver epoxy to connect these four actuators in series. The haptic device was mounted on suspended flexible foam tape and connected to a waveform generator (33210A, Keysight) via an amplifier (WMA-300, Falco Systems). Out-of-plane displacement was recorded using a laser Doppler vibrometer (OFV-5000, Polytec). A computer-controlled x-y stage moved the device for 2D mapping. The entire setup was controlled through a LabVIEW program, as demonstrated in a previous paper.

Dielectric properties measurement: Relative permittivity $\varepsilon_r$ and dielectric losses tanδ of fabricated actuators were measured as functions of DC voltage with a probing AC signal of 0.05V at 1 kHz, using TF Analyzer 2000 (aixACCT).

**References**

1   Shi, Y. & Shen, G. Haptic Sensing and Feedback Techniques toward Virtual Reality. Research **7**, 0333 (2024).

2	Yu, X. et al. Skin-integrated wireless haptic interfaces for virtual and augmented reality. Nature **575**, 473-479 (2019).

3	Tang, Y. et al. Advancing haptic interfaces for immersive experiences in the metaverse. Device **2** (2024).

4	Choi, C. et al. Surface haptic rendering of virtual shapes through change in surface temperature. Sci. Robot. **7**, eabl4543 (2022).

5	Gao, Y. et al. Advances in materials for haptic skin electronics. Matter **7**, 2826-2845 (2024).

6	Chen, J., Teo, E. H. T. & Yao, K. in Actuators.  104 (MDPI).

7	Biet, M., Giraud, F. & Lemaire-Semail, B. Squeeze film effect for the design of an ultrasonic tactile plate. IEEE Trans. Ultrason. Ferroelectr. Freq. Control **54**, 2678-2688 (2007).

8	Wiertlewski, M., Fenton Friesen, R. & Colgate, J. E. Partial squeeze film levitation modulates fingertip friction. Proc. Natl. Acad. Sci. U.S.A. **113**, 9210-9215 (2016).

9	Giraud, F. & Giraud-Audine, C. Piezoelectric Actuators: Vector Control Method: Basic, Modeling and Mechatronic Design of Ultrasonic Devices.  (Butterworth-Heinemann, 2019).

10	Biet, M., Casiez, G., Giraud, F. & Lemaire-Semail, B. in 2008 symposium on haptic interfaces for virtual environment and teleoperator systems.  41-48 (IEEE).

11	Glinsek, S. et al. Fully transparent friction‐modulation haptic device based on piezoelectric thin film. Adv. Funct. Mater. **30**, 2003539 (2020).

12	Glinsek, S. et al. Inkjet‐printed piezoelectric thin films for transparent haptics. Adv. Mater. Technol. **7**, 2200147 (2022).

13	Hua, H., Chen, Y., Tao, Y., Qi, D. & Li, Y. A highly transparent haptic device with an extremely low driving voltage based on piezoelectric PZT films on glass. Sensor Actuat. A-Phys. **335**, 113396 (2022).

14	Song, L. et al. Crystallization of piezoceramic films on glass via flash lamp annealing. Nature Communications **15**, 1890 (2024).

15	Huangfu, G. et al. Giant electric field–induced strain in lead-free piezoceramics. Science **378**, 1125-1130 (2022).

16	Wu, J., Xiao, D. & Zhu, J. Potassium–sodium niobate lead-free piezoelectric materials: past, present, and future of phase boundaries. Chem. Rev. **115**, 2559-2595 (2015).

17	Saito, Y. et al. Lead-free piezoceramics. Nature **432**, 84-87 (2004).

18  Kuentz, H. et al. KNN lead-free technology on 200 mm Si wafer for piezoelectric actuator applications. Sensor Actuat. A-Phys. **372**, 115370 (2024).

19  Shrout, T. R. & Zhang, S. J. Lead-free piezoelectric ceramics: Alternatives for PZT? J. Electroceram. **19**, 113-126 (2007).

20  Liu, Y. X. et al. Multi-Length Engineering of (K, Na) NbO3 Films for Lead-Free Piezoelectric Acoustic Sensors with High Sensitivity. Adv. Funct. Mater. **34**, 2312699 (2024).

21  Waqar, M. et al. Large Electromechanical Response in a Polycrystalline Alkali-Deficient (K, Na) NbO3 Thin Film on Silicon. Nano Lett. **23**, 11026-11033 (2023).

22  Song, L. et al. Piezoelectric thick film for power-efficient haptic actuator. Appl. Phys. Lett. **121** (2022).

23  More information of SCHOOT AF32 glass is available on : https://www.pgo-online.com/intl/af32.html.

24  Kovacova, V. et al. Comparative solution synthesis of Mn doped (Na, K) NbO3 thin films. Chem. Eur. J. **26**, 9356-9364 (2020).

25  Jacques, L., Kovacova, V., Yang, J. I. & Trolier-McKinstry, S. Activation energies for crystallization of manganese-doped (K, Na) NbO3 thin films deposited from a chemical solution. J. Am. Ceram. Soc. **104**, 4968-4976 (2021).

26  Alluri, N. R. et al. Enhanced Electromechanical Properties of Solution-Processed K0.5Na0.5NbO3 Thin Films. arXiv:2502.21066 (2025).

27  Shibata, K., Watanabe, K., Kuroda, T. & Osada, T. KNN lead-free piezoelectric films grown by sputtering. Applied Physics Letters **121**, doi:10.1063/5.0104583 (2022).

28  Ueda, K., Kweon, S.-H., Hida, H., Mukouyama, Y. & Kanno, I. Transparent piezoelectric thin-film devices: Pb (Zr, Ti) O3 thin films on glass substrates. Sensor Actuat. A-Phys. **327**, 112786 (2021).

29  ICDD database PDF4+ v.19. (2019).

**Data and materials availability**

All data is available in the main text or the Supplementary Information. Correspondence and requests for materials should be addressed to E.D. Source data are provided with this paper.

**Acknowledgement**

Luxembourg National Research Fund (FNR) is acknowledged for the financial support through project FLASHPOX (C21/MS/16215707).

**Author contributions**

E.D. suggested the experimental study. S.G., N.R.A., V.K., and J.C. developed the KNN solution. S.G., N.R.A., and V.K. developed the deposition process for KNN thin films. L.S. designed, fabricated, and characterised the transparent haptic devices. L.S. also performed the finite element modeling. B.M. conducted the SEM analysis. M.M. deposited FTO and ITO on glass. A.B.M. developed the post-annealing recipes for FTO. L.S. wrote the manuscript with E.D. and S.G. All authors contributed to the final version of the manuscript. S.G. obtained the funding.

**Competing interests**

The authors declare no competing interests.

# Supplementary information

**Highly Transparent Lead-Free Piezoelectric Haptic Device**

Longfei Song,[1,2,4] Sebastjan Glinsek,[1,2] Nagamalleswara Rao Alluri,[1,2] Veronika Kovacova,[1,2] Michele Melchiorr,[3] Alfredo Blazquez Martinez,[1,2] Barnik Mandal,[1,2] Juliette Cardoletti,[1,2] Emmanuel Defay[1,2*]

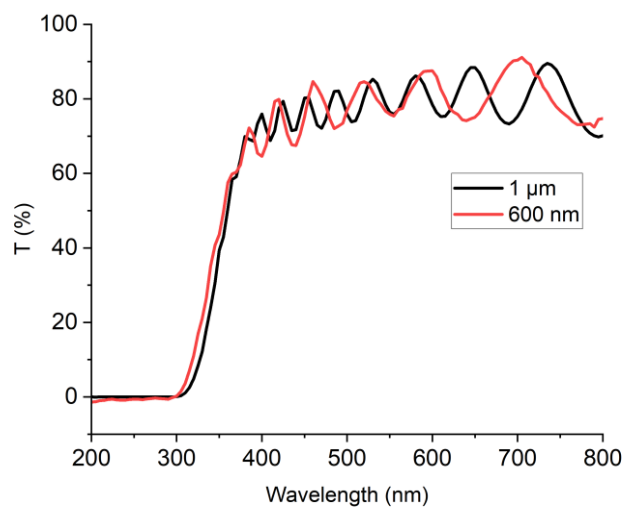

Figure S1. Transmittance of 1μm and 600nm thick KNN film deposited on FTO glass.

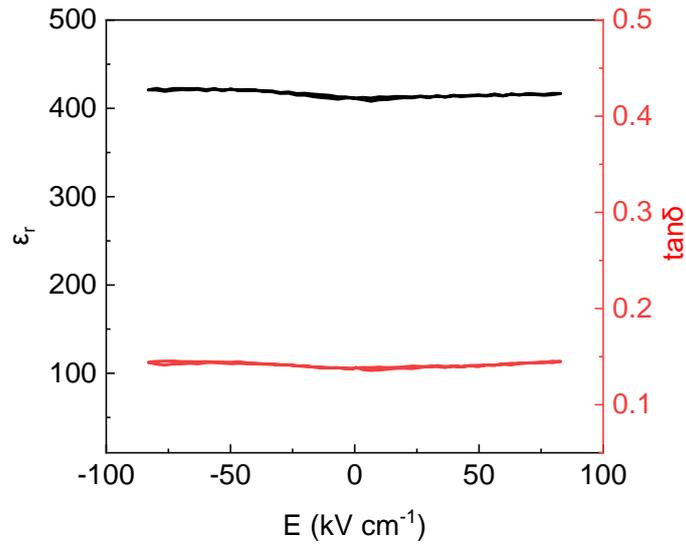

Fig. S2 Dielectric properties of the ITO/KNN/FTO actuator. Relative permittivity $\varepsilon_r$ and dielectric losses $\tan\delta$ as function of electric field, measured by applying a staircase-like DC bias with an AC signal of 0.05 V at 1 kHz using TF Analyzer 2000 (aixACCT).

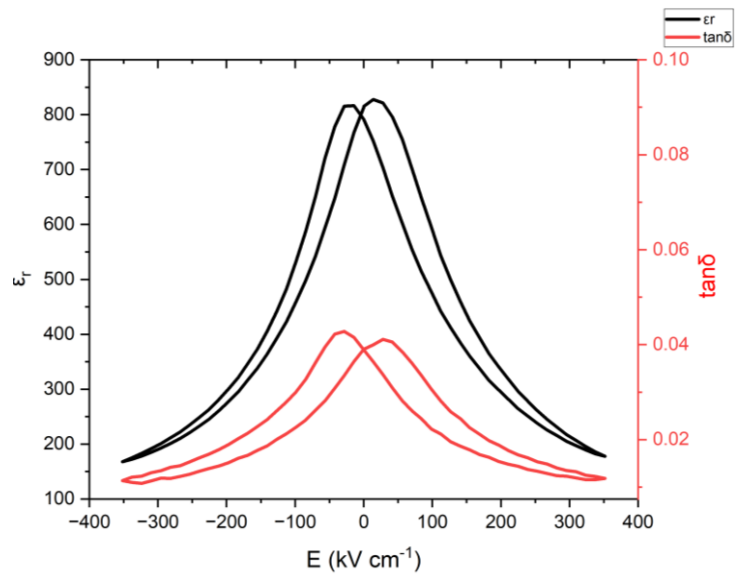

Fig. S3 Dielectric properties of the 400 nm-thick KNN film on platinised silicon substrate (Pt/KNN/Pt/Si), measured by applying a staircase-like DC bias with an AC signal of 0.05 V at 1 kHz using TF Analyzer 2000 (aixACCT).

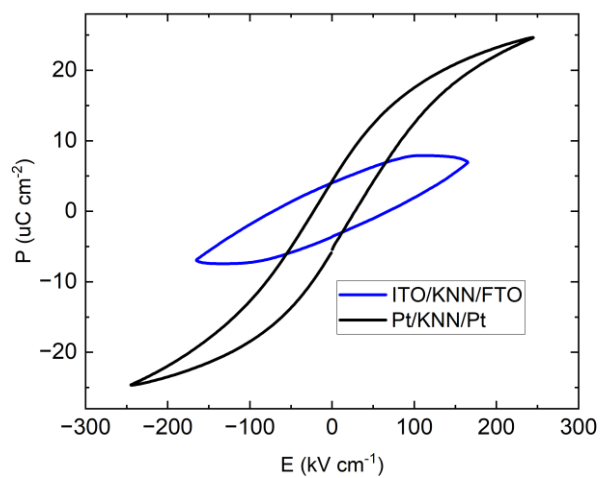

Fig S4. Polarisation versus electric field (P-E) hysteresis loops of KNN films deposited on platinised silicon (Pt/KNN/Pt/Si device), and deposited on FTO glass (ITO/KNN/FTO/glass), respectively. The P-E loop was measured with a triangular waveform at 100 Hz using TF Analyzer 2000 (aixACCT).